\documentclass[preprint2]{aastex}

\usepackage{CJK}
\usepackage{psfig}
\usepackage{amsmath}

\newcommand{\E}[1]{\times 10^{#1}}
      \newcommand{\ps}{\,{\rm s}$^{-1}$}
    \newcommand{\Msun}{{\rm M}_{\rm \odot}}
    \newcommand{\km}{\,{\rm km}}

\newcommand{\ncol}{$N({\rm H_2})$}
\newcommand{\xray}{X-ray}

\newcommand{\du}{$d_{1.95}$} 

\newcommand{\vlsr}{$V_{\rm LSR}$}       \newcommand{\tmb}{$T_{\rm mb}$}
\newcommand{\twCO}{$^{12}$CO}  \newcommand{\thCO}{$^{13}$CO}
\newcommand{\CeiO}{C$^{18}$O} 
\newcommand{\snr}{{HB~3}}

\shorttitle{SNR HB~3 and W3 Region}
\shortauthors{Zhou et al.}

\begin{document}
\begin{CJK*}{UTF8}{gbsn}

\title{Interaction between the Supernova Remnant HB~3 and the Nearby Star-Forming Region W3}

\author{Xin Zhou (周鑫)\altaffilmark{1,2}, Ji Yang (杨戟)\altaffilmark{1,2}, Min Fang (房敏)\altaffilmark{1,2}, Yang Su (苏扬)\altaffilmark{1,2}, Yan Sun (孙燕)\altaffilmark{1,2}, Yang Chen (陈阳)\altaffilmark{3,4}}
\affil{$^1$Purple Mountain Observatory, Chinese Academy of Sciences, 2 West Beijing Road, Nanjing 210008, China; xinzhou@pmo.ac.cn \\
$^2$Key Laboratory of Radio Astronomy, Chinese Academy of Sciences, Nanjing 210008, China \\
$^3$Department of Astronomy, Nanjing University, 163 Xianlin Avenue, Nanjing 210023, China \\
$^4$Key Laboratory of Modern Astronomy and Astrophysics, Nanjing University, Ministry of Education, Nanjing 210093, China
}


\begin{abstract}
We performed millimeter observations in CO lines toward the supernova remnant (SNR) \snr. 
Substantial molecular gas around $-45$~\km\ps\ is detected in the conjunction region between the SNR~\snr\ and the nearby W3 complex. 
This molecular gas is distributed along the radio continuum shell of the remnant.
Furthermore, the shocked molecular gas indicated by line wing broadening features is also distributed along the radio shell and inside it.
By both morphological correspondence and dynamical evidence, we confirm that the SNR~\snr\ is interacting with the $-45$~\km\ps\ molecular cloud (MC), in essence, with the nearby H~{\sc ii} region/MC complex W3.
The red-shifted line wing broadening features indicate that the remnant is located at the nearside of the MC.
With this association, we could place the remnant at the same distance as the W3/W4 complex, which is $1.95\pm0.04$~kpc.
The spatial distribution of aggregated young stellar object candidates (YSOc) shows a correlation to the shocked molecular strip associated with the remnant.
We also find a binary clump of CO at ($l=132^{\circ}.94, b=1^{\circ}.12$) around $-51.5$~\km\ps\ inside the projected extent of the remnant, and it is associated with significant mid-infrared (mid-IR) emission. 
The binary system also has a tail structure resembling the tidal tails of interacting galaxies.
According to the analysis of CO emission lines, the larger clump in this binary system is about stable, and the smaller clump is significantly disturbed.
\end{abstract}


\section{Introduction}
The supernova remnant (SNR) \snr\ was discovered in the radio band \citep{BrownHazard1953,Williams+1966,Caswell1967}, and progressively, multi-wavelength emissions were detected from it.
It has an angular size of $90'\times120'$ and a radio spectral index of $-0.56$ \citep{Landecker+1987,Fesen+1995,Reich+2003,TianLeahy2005,Green2007}.
\snr\ is considered to be an evolved SNR, as indicated by a strong radio-optical correlation plus a multishell structure \citep{Fesen+1995}.
Characterized by shell-like radio continuum morphology and centrally peaked thermal X-ray emission, 
\snr\ is identified as a mixed-morphology (MM) or thermal composite SNR \citep[][and references therein]{LazendicSlane2006}.
By spectral analysis to {\it ASCA} and {\it XMM-Newton} \xray\ observations, \cite{LazendicSlane2006} derived the velocity of the remnant's blast wave $340\pm{37}$~\km\ps, the remnant's ambient particle density $0.32\pm0.10$~cm$^{-3}$, the remnant's age ($3.00\pm0.33$)$\E{4}$~yr, and the explosion energy ($3.4\pm1.5$)$\E{50}$~ergs.

The remnant is adjacent in the sky to the H~{\sc ii} region/molecular cloud (MC) complex W3 with a potential association between them \citep{Landecker+1987}.
\cite{Routledge+1991} examined the H~{\sc i} and \twCO~(J=1--0) line emissions, and found a bright \twCO\ ``bar" near $-43$~\km\ps\ that is morphologically corresponding to \snr's enhanced radio continuum emission, which supports the association between the remnant and the W3 complex. 
The distance of the W3 complex is $1.95\pm0.04$~kpc, which was determined by triangulation method \citep{Xu+2006}.
A large H~{\sc i} shell surrounds the remnant was found in velocity from $-25$ to $-43$~\km\ps\ \citep{Routledge+1991, Normandeau+1997}.
No OH 1720~MHz maser was found to be associated with the shock of \snr\ \citep{Koralesky+1998}.
Broadened \twCO~(J=2--1) line emission was detected toward the north of \snr, which was confirmed to be associated with the H~{\sc ii} region W3~(OH) but not \snr\ \citep[][and references therein]{Kilpatrick+2016}.

In this paper, we present CO line observations fully covering the SNR~\snr, and confirm that \snr\ is interacting with the nearby W3 complex.
For convenience, we introduce the factor of distance \du\ stands for $d$/(1.95~kpc), where $d$ is the distance to \snr.
The observations are described in Sections~\ref{sec:obs}. In Section~\ref{sec:result} and Section~\ref{sec:discuss}, we present the results and physical interpretations, respectively. The conclusions are summarized in Section~\ref{sec:conclusion}.

\section{Observations}\label{sec:obs}
The observations of \twCO, \thCO, and \CeiO\ line emissions toward the SNR~\snr\
were made from January 2012 to February 2014 with the Purple Mountain Observatory Delingha (PMODLH) 13.7~m millimeter-wavelength telescope \citep{Zuo+2011},
which is a part of the Milky Way Image Scroll Painting (MWISP)--CO line survey project\footnote{http://www.radioast.nsdc.cn/yhhjindex.php}.
The three CO lines were observed simultaneously with the $3\times3$ multibeam sideband separation superconducting receiver \citep{Shan+2012}.
A Fast Fourier Transform (FFT) spectrometer with a total bandwidth of 1000 MHz and 16384 channels was used as a back end.
The typical system temperatures were around 230 K for \twCO\ and around 140 K for \thCO\ and \CeiO, and the variations among different beams were less than $15\%$. 
The total error in pointing and tracking was within $5''$. The half-power beam width (HPBW) was about $51''$.
The main-beam efficiencies $\eta_{\rm mb}$ were $\sim$44\% for USB and $\sim$48\% for LSB, and the differences among the beams were less than 8\%.
These parameters were obtained by using the five-point pointing observations toward known or calibrator sources, and the standard chopper-wheel method was used for temperature calibration (\citealt{UlichHaas1976}; see details in Status Report on the 13.7~m Millimeter-Wave Telescope\footnote{Status Report on the 13.7 m Millimeter-Wave Telescope for each observing season is available at http://www.radioast.csdb.cn/zhuangtaibaogao.php}).
The spectral resolutions were 0.17~\km\ps\ for \twCO~(J=1--0) and 0.16~\km\ps\ for both \thCO~(J=1--0) and \CeiO~(J=1--0).
We mapped a $210'\times 210'$ area that contains the full extent of the SNR~\snr\ via on-the-fly (OTF) observing mode, and the data was meshed with a grid spacing of $30''$.
Using the emission-free velocity ranges of $-200$ to $-115$~\km\ps\ and $100$ to $150$~\km\ps, we performed a linear baseline subtraction, and got the average RMS noises of all final spectra of about 0.5~K for \twCO~(J=1--0) in $0.17$~\km\ps\ channel and about 0.3~K for \thCO~(J=1--0) and \CeiO~(J=1--0) in $0.16$~\km\ps\ channels.
All data were reduced using the GILDAS/CLASS package\footnote{http://www.iram.fr/IRAMFR/GILDAS}.

408~MHz radio continuum emission data were obtained from the Canadian Galactic Plane Survey \citep[CGPS;][]{Taylor+2003}.
Near-infrared (near-IR) $J~H~K_{\rm S}$ data of the Two-Micron All Sky Survey \citep[2MASS, ][]{2MASS2006} were obtained, of which the 10$\sigma$ point-source detection levels are better than 15.8, 15.1, and 14.3~mag, respectively.
IR photometric data of the survey of Wide-field Infrared Survey Explorer \citep[WISE,][]{WISE2010} were also obtained. 
The angular resolutions are 6$''$.1, 6$''$.4, 6$''$.5, and 12$''$.0\ in the four WISE bands (3.4, 4.6, 12, and 22~$\mu$m), respectively, and the achieved 5$\sigma$ point source sensitivities are better than 0.08, 0.11, 1, and 6 mJy in the four bands, respectively.
The 2MASS $H~K_{\rm s}$ bands and the first three WISE bands data were used for point source analyses, with the photometric error of the selected point source less than 0.1 mag for the 2MASS data and the signal-to-noise ratio greater than 10 for the WISE data.

\section{Results}\label{sec:result}
\subsection{Morphology}\label{subsec:morphology}

\begin{figure*}[ptbh!]
\centerline{{\hfil\hfil
\psfig{figure=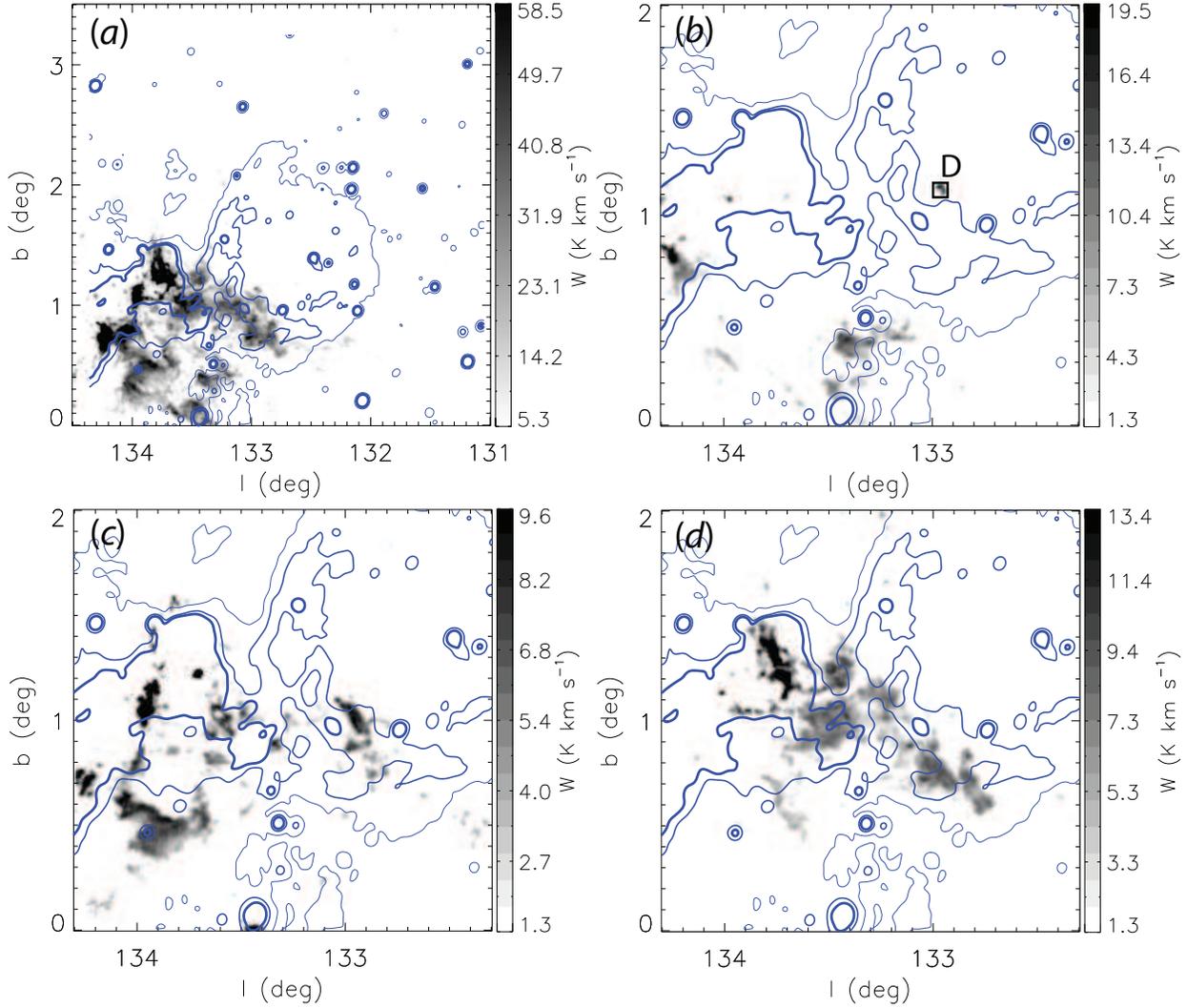,height=5.5in,angle=0, clip=}
\hfil\hfil}}
\caption{\twCO~(J=1--0) emission maps integrated over four velocity ranges, (a) $-53$ to $-35$~\km\ps, (b) $-52$ to $-51$~\km\ps, (c) $-46$ to $-45$~\km\ps, and (d) $-42$ to $-41$~\km\ps. 
We show the integrated CO emission in the whole observed region in panel ({\it a}), and the representative channel maps (with the velocity width of 1~\km\ps and the RMS of $\sim$$0.21$~K) in the other panels.
The minimum value of each map is $5\sigma$.
The contours repeated on each panel are the CGPS radio continuum emission at 408~MHz, with the thicker lines indicating the higher intensities. The contour levels are 65, 87.5, and 110~K. 
The region that contains a binary clump is marked with a black box around ($l=132^{\circ}.94, b=1^{\circ}.12$) in panel ({\it b}), namely region~D.
}
\label{f:stamp12}
\end{figure*}

The \twCO\ intensity maps shown in Figure~\ref{f:stamp12} exhibit the distribution of molecular gas around $-45$~\km\ps. 
There is substantial molecular gas associated with the nearby H~{\sc ii} complex, i.e.\ W3, \cite[$l=133^{\circ}.8, b=1^{\circ}.2$; see][]{Green2007}, which is present in the whole velocity range of $-53$ to $-35$~\km\ps. 
There is also molecular gas in the remnant region, distributed along the radio continuum shell of the remnant. 
Based on their spatial and velocity continuities, we confirm that these molecular gases are from a same MC.
In the middle of the conjunction area between W3 and \snr, \twCO~(J=1--0) emission protrudes from the W3 region into the \snr\ region.
There is no strong CO emission detected in the north and northwest of the remnant.
\cite{Kilpatrick+2016} detected \twCO~(J=2-1) emission toward two small regions ($10'\times30'$ and $10'\times10'$) in the far north of the remnant, however, they suggested that the detected broad molecular line region is associated with the W3 complex, but not the remnant.
The distribution of molecular gas is morphologically consistent with the non-thermal radio continuum emission of the remnant, which shows blowout morphology from the north to northwest and bright shell from the east to southwest.
It is somewhat similar to the blowout morphology in the SNR~N132D \citep[e.g.,][]{DickelMilne1995, XiaoChen2008}, which was suggested to be shaped by the shock impacting on a stellar wind-bubble shell \citep{Hughes1987, Chen+2003}.
The radio continuum emissions of the remnant and the W3 complex are in conjunction with each other as well (around $l=133^{\circ}.2, b=1^{\circ}.0$), without a clear border between them.
Along with the velocity component around -45\km\ps, there are also the other velocity components, around 0~\km\ps, $-75$~\km\ps, and $-105$~\km\ps.
We have checked them, and find no morphological correlations to the remnant.

\begin{figure*}[ptbh!]
\centerline{{\hfil\hfil
\psfig{figure=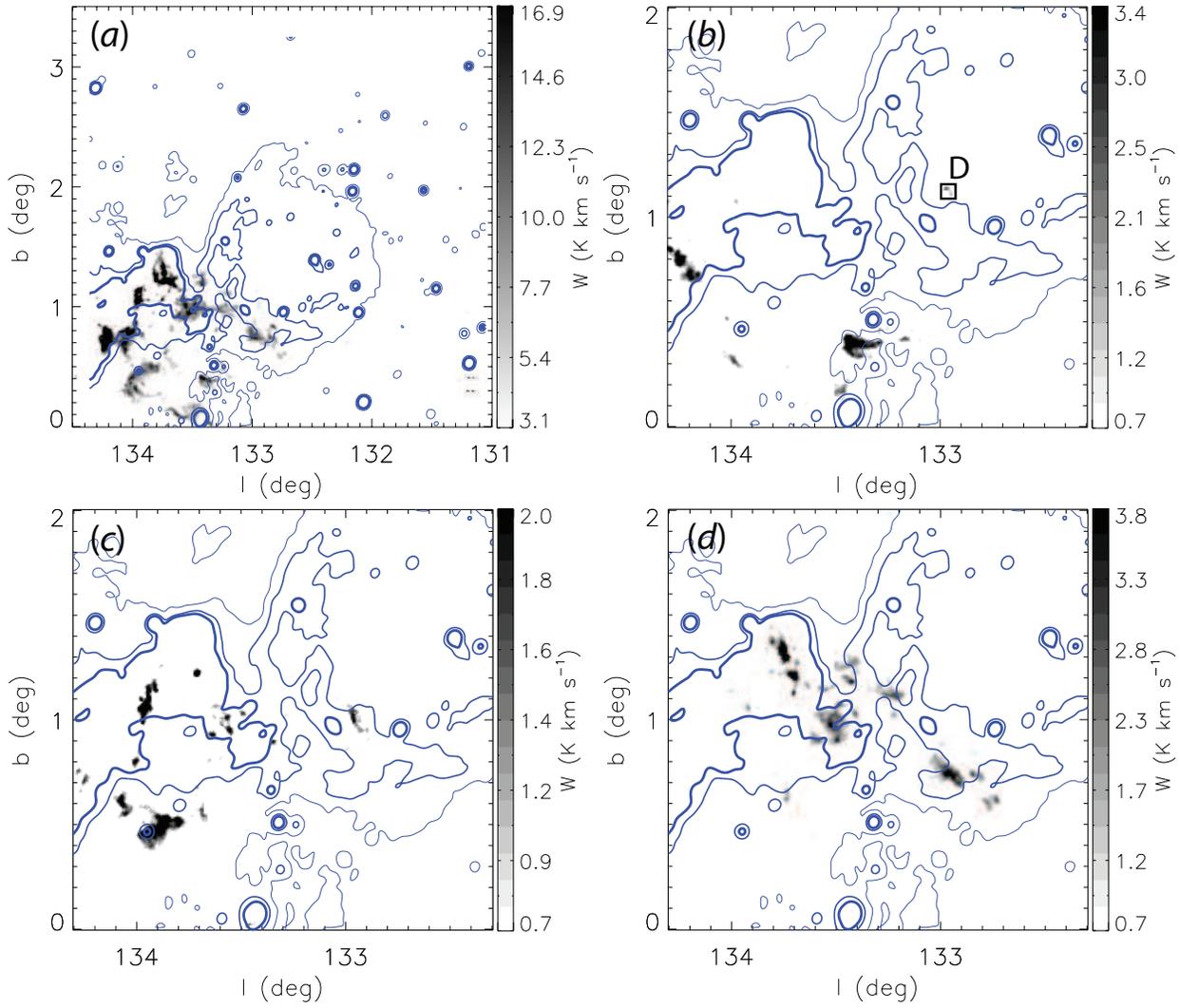,height=5.5in,angle=0, clip=}
\hfil\hfil}}
\caption{
The same as Figure~\ref{f:stamp12}, but for \thCO~(J=1--0) emission. The RMS of 1~\km\ps\ channel map is $\sim$$0.12$~K.
}
\label{f:stamp13}
\end{figure*}

Figure~\ref{f:stamp13} shows the \thCO\ intensity maps produced in the same manner as Figure~\ref{f:stamp12}. 
The \thCO\ emission is less extended than \twCO, with the peak in the W3 region.
We also detect some \thCO\ emissions along the radio shell of the remnant.
There is no significant \CeiO\ emission detected in the remnant region. 

\begin{figure*}[ptbh!]
\centerline{{\hfil\hfil
\psfig{figure=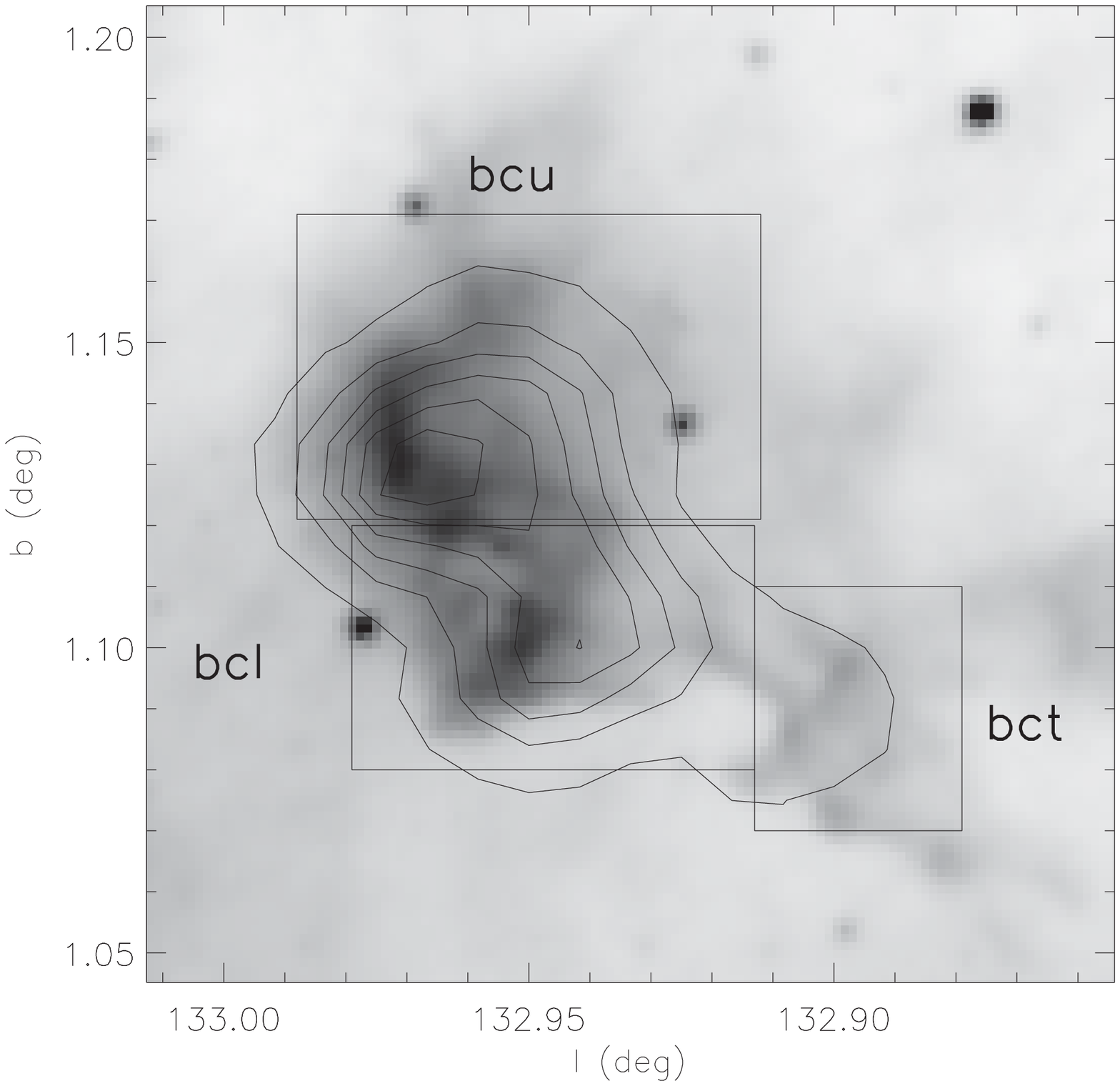,height=5.8in,angle=0, clip=}
\hfil\hfil}}
\caption{
WISE 12 $\mu$m image of the binary clump overlaid with integrated intensity contours of \twCO~(J=1--0) emission in the velocity range of $-53$ to $-49$~\km\ps. The contours have the minimum of $5\sigma$ with a step of $10\sigma$. The RMS is $\sim$0.42~K.
The binary clump is divided into three regions, namely, the binary clump upper region, the binary clump lower region, and the binary clump tail region, which are indicated by rectangles labeled bcu, bcl, and bct, respectively.
}
\label{f:bcir}
\end{figure*}

\begin{figure*}[ptbh!]
\centerline{{\hfil\hfil
\psfig{figure=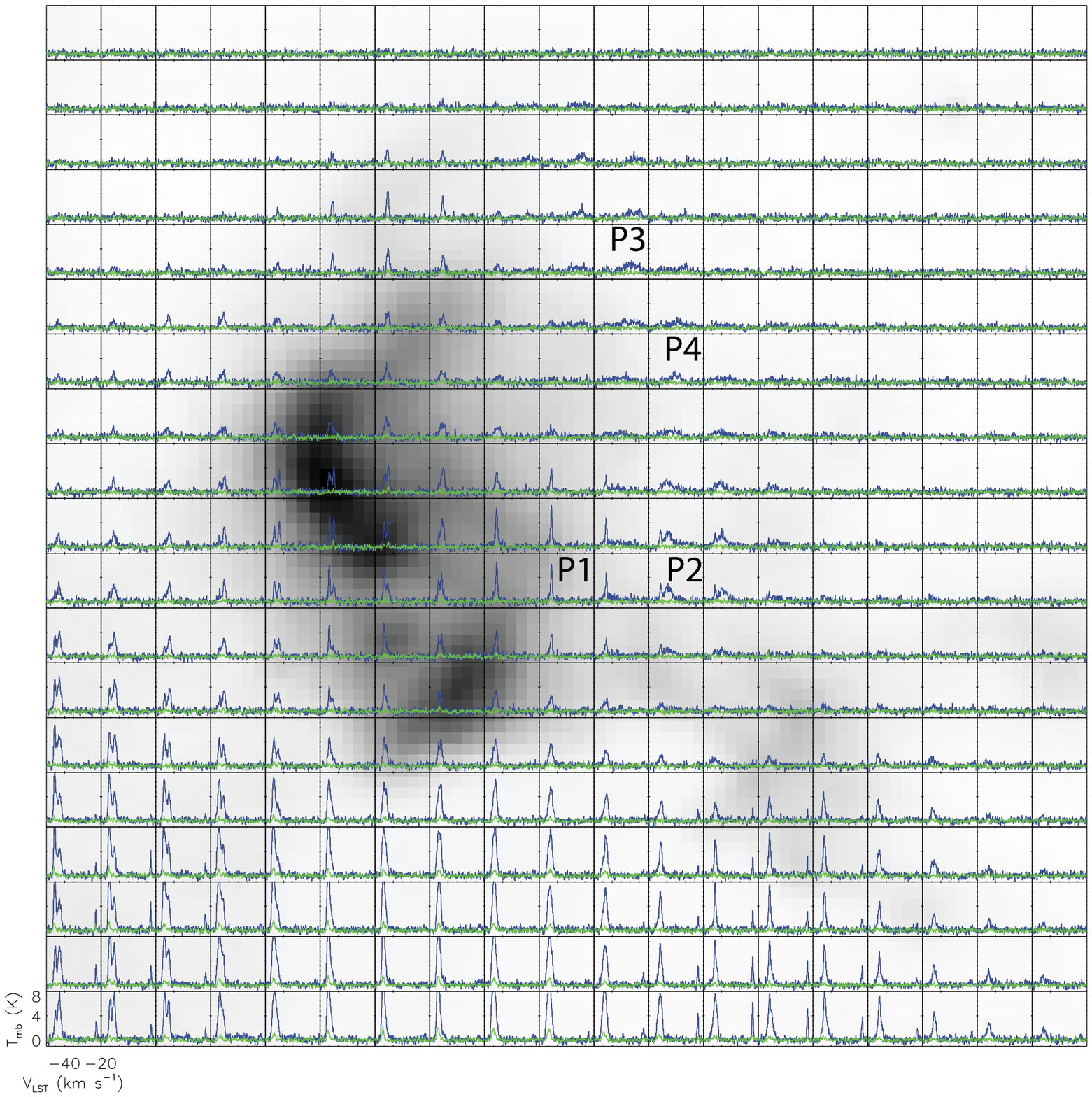,height=6.0in,angle=0, clip=}
\hfil\hfil}}
\caption{
Grid of \twCO~(J=1--0) (blue) and \thCO~(J=1--0) (green) spectra around $-45$~\km\ps\ in region~D (see panel ({\it b}) in Figures~\ref{f:stamp12} and \ref{f:stamp13}), superposed on the WISE 22~$\mu$m image. The image covers the same region shown in Figure~\ref{f:bcir}. The grid spacing is 30$''$, corresponding to the length of 0.28\du~pc. 
Four representative points are selected for spectral analysis, and labeled with P1, P2, P3, and P4, respectively.
}
\label{f:regdspec}
\end{figure*}

Particularly, a binary clump is found at ($l=132^{\circ}.94, b=1^{\circ}.12$) around $-51.5$~\km\ps\ inside the remnant's radio shell, which is prominent in both \twCO\ and \thCO\ emissions. 
This binary clump, denoted as region~D, is indicated in panel ({\it b}) in Figures~\ref{f:stamp12} and \ref{f:stamp13}.
The two clumps in the system have a small velocity shift of $\sim$0.3~\km\ps.
This $-51.5$~\km\ps\ binary clump has a tail extending toward the western region too (see Figure~\ref{f:bcir}), which is a sign of interaction between the two clumps.
In the same region, near the binary clump, there is also a small protrusion from the $-45$~\km\ps\ MC located around the radio shell of the remnant into the inner region of the remnant (see panel ({\it c}) in Figure~\ref{f:stamp12}). 

Significant mid-IR emissions (12~$\mu$m and 22~$\mu$m) are also detected in the south-eastern part in region~D (see Figures~\ref{f:bcir} and \ref{f:regdspec}).
The mid-IR emission has two peaks similar to the $-51.5$~\km\ps\ component, but with the position shifted to the $-45$~\km\ps\ protrusion. 
In addition, the mid-IR emission has a tail structure corresponding to that of the $-51.5$~\km\ps\ component, where no significant $-45$~\km\ps\ emission present.
There are also weak mid-IR emissions beyond the binary clump in the west, northwest, and north (see Figure~\ref{f:bcir}), with associated \twCO~(J=1--0) emission around $-45$~\km\ps\ but not $-51.5$~\km\ps.

\subsection{Dynamics}\label{sec:dynamics}
\begin{figure*}[ptbh!]
\centerline{{\hfil\hfil
\psfig{figure=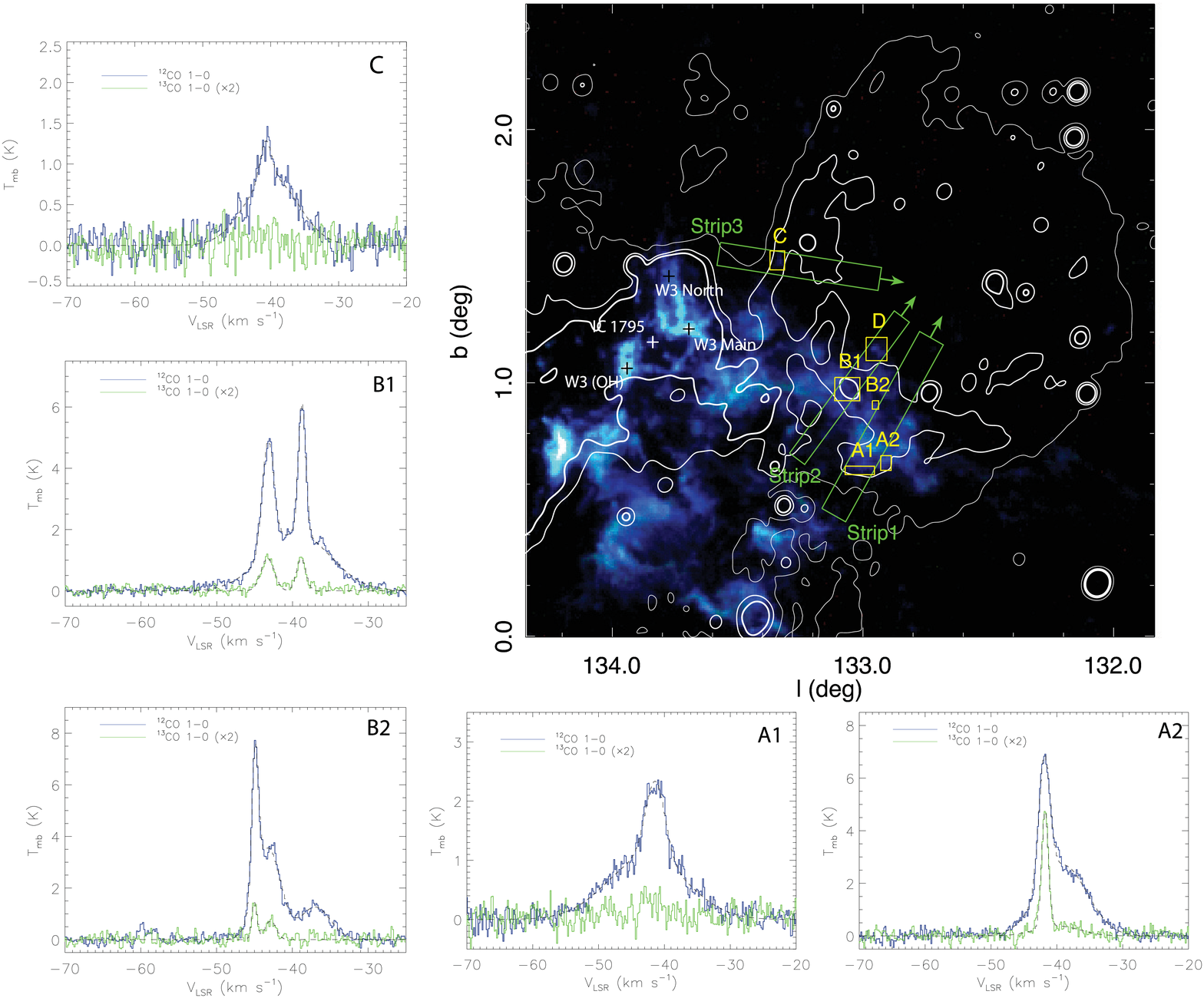,height=5.2in,angle=0, clip=}
\hfil\hfil}}
\caption{
Right top panel: integrated intensity map of \twCO~(J=1--0) emission (blue), \thCO~(J=1--0) (green), and \CeiO~(J=1--0) (red) in the velocity range of $-60$ to $-30$~\km\ps, overlaid with the same 408~MHz continuum contours as in Figure~\ref{f:stamp12}.
The green and yellow regions are selected for generating position-velocity maps and for spectral analysis, respectively. Region~D is the same as that in Figures~\ref{f:stamp12} and \ref{f:stamp13}.
The prominent features of the W3 complex are labeled.
Other five panels: the \twCO~(J=1--0) (blue) and \thCO~(J=1--0) (green) spectra extracted from the corresponding regions A1, A2, B1, B2, and C, respectively. The \thCO~(J=1--0) spectra are multiplied by a factor of 2 for a better visibility. 
The spectra are fitted by multiple Gaussian components, with the fitting results shown by black dashed lines (see the fitted parameters in Table~\ref{tab:fitpara}).
}
\label{f:tricolor}
\end{figure*}

\begin{table*}\footnotesize
\begin{center}
\caption{Fitted parameters of thermal molecular lines.\label{tab:fitpara}}
\begin{tabular}{llcccccc}
\tableline\tableline
region&component&Line&Peak \tmb\tablenotemark{[1]}&Center \vlsr\tablenotemark{[2]}&FWHM\\
&&&(K)&(\km\ps)&(\km\ps)\\
\tableline
A1&narrow&\twCO~(J=1--0)&1.31$\pm{0.06}$&-41.38$\pm{0.06}$&3.5$\pm{0.2}$\\
&&\thCO~(J=1--0)&$\leq{0.08}$\tablenotemark{[3]}&-&-\\
&broad&\twCO~(J=1--0)&1.05$\pm{0.05}$&-42.7$\pm{0.2}$&14.1$\pm{0.5}$\\
&&\thCO~(J=1--0)&$\leq{0.08}$\tablenotemark{[3]}&-&-\\
\tableline
A2&narrow&\twCO~(J=1--0)&4.83$\pm{0.08}$&-41.89$\pm{0.02}$&2.01$\pm{0.04}$\\
&&\thCO~(J=1--0)&2.21$\pm{0.05}$&-41.79$\pm{0.02}$&1.22$\pm{0.04}$\\
&broad&\twCO~(J=1--0)&2.70$\pm{0.05}$&-39.10$\pm{0.07}$&8.7$\pm{0.2}$\\
&&\thCO~(J=1--0)&0.19$\pm{0.1}$&-39.9$\pm{0.5}$&7.9$\pm{0.9}$\\
\tableline
B1&narrow1&\twCO~(J=1--0)&3.56$\pm{0.06}$&-43.23$\pm{0.02}$&1.80$\pm{0.04}$\\
&&\thCO~(J=1--0)&0.53$\pm{0.03}$&-43.29$\pm{0.05}$&2.1$\pm{0.1}$\\
&narrow2&\twCO~(J=1--0)&4.26$\pm{0.07}$&-38.744$\pm{0.009}$&1.13$\pm{0.03}$\\
&&\thCO~(J=1--0)&0.52$\pm{0.03}$&-38.87$\pm{0.03}$&1.47$\pm{0.09}$\\
&broad&\twCO~(J=1--0)&1.89$\pm{0.04}$&-39.48$\pm{0.08}$&10.4$\pm{0.2}$\\
&&\thCO~(J=1--0)&$\leq{0.05}$\tablenotemark{[3]}&-&-\\
\tableline
B2&narrow1&\twCO~(J=1--0)&5.6$\pm{0.2}$&-44.92$\pm{0.01}$&1.02$\pm{0.03}$\\
&&\thCO~(J=1--0)&0.71$\pm{0.06}$&-44.99$\pm{0.04}$&0.82$\pm{0.08}$\\
&narrow2&\twCO~(J=1--0)&3.54$\pm{0.06}$&-43.19$\pm{0.06}$&3.9$\pm{0.1}$\\
&&\thCO~(J=1--0)&0.36$\pm{0.04}$&-42.9$\pm{0.1}$&1.7$\pm{0.3}$\\
&broad&\twCO~(J=1--0)&1.26$\pm{0.05}$&-37.0$\pm{0.1}$&4.7$\pm{0.3}$\\
&&\thCO~(J=1--0)&$\leq{0.08}$\tablenotemark{[3]}&-&-\\
\tableline
C&narrow&\twCO~(J=1--0)&0.47$\pm{0.07}$&-40.7$\pm{0.2}$&1.9$\pm{0.4}$\\
&&\thCO~(J=1--0)&$\leq{0.08}$\tablenotemark{[3]}&-&-\\
&broad&\twCO~(J=1--0)&0.83$\pm{0.05}$&-39.9$\pm{0.2}$&9.3$\pm{0.5}$\\
&&\thCO~(J=1--0)&$\leq{0.08}$\tablenotemark{[3]}&-&-\\
\tableline\tableline
\end{tabular}
\tablenotetext{[1]}{\tmb\ is the brightness temperature, and is corrected for beam efficiency using \tmb=$T_{\rm A}^{*}/\eta_{\rm mb}$.}
\tablenotetext{[2]}{\vlsr\ is the velocity with respect to the local standard of rest.}
\tablenotetext{[3]}{No \thCO~(J=1--0) emission visible, where we use the value of RMS as an upper limit.}
\end{center}
\end{table*}

Broad \twCO~(J=1--0) emission lines are detected in many places, and they have different velocities at different locations. In Figure~\ref{f:tricolor}, the spectra extracted from five selected regions are shown, namely A1, A2, B1, B2, and C. There are also broad \twCO~(J=1--0) emission lines in region~D (see Figure~\ref{f:regdspec}).
These regions are distributed along the remnant's radio shell around the conjunction area between the remnant and the W3 complex as well inside it.
We do not find any evidence of association between these broad \twCO~(J=1--0) emission-line regions and the H~{\sc ii} regions in the W3 complex.
According to the correspondence of spatial distribution and as an exclusive source of disturbance, the broadened emission lines are originated from the molecular gas impacted by the remnant shock, which confirms the association between \snr\ and the MCs in the W3 complex.
We have performed Gaussian fitting to the emission lines, and the fitted parameters are listed in Table~\ref{tab:fitpara}.
For most of the broad \twCO~(J=1--0) components, the line centers are red-shifted comparing to the narrow components except in region~A1. 
It indicates that the remnant is at the nearside of the MCs.
Note that, in region~C, both the narrow and broad components are weak, and the intensity of the broad component is stronger than that of the narrow component, indicating that there is not much quiet molecular gas left in this region.

\begin{table*}\footnotesize
\begin{center}
\caption{Fitted parameters of thermal molecular lines in region~D.\label{tab:fitparad}}
\begin{tabular}{llcccccc}
\tableline\tableline
region \&point&component&Line&Peak \tmb &Center \vlsr&FWHM\\
&&&(K)&(\km\ps)&(\km\ps)\\
\tableline
whole&narrow1&\twCO~(J=1--0)&5.45$\pm{0.06}$&-51.615$\pm{0.006}$&1.02$\pm{0.02}$\\
&&\thCO~(J=1--0)&0.80$\pm{0.04}$&-51.81$\pm{0.02}$&0.84$\pm{0.05}$\\
&narrow2&\twCO~(J=1--0)&0.94$\pm{0.07}$&-45.17$\pm{0.04}$&0.89$\pm{0.09}$\\
&&\thCO~(J=1--0)&0.16$\pm{0.05}$&-45.02$\pm{0.08}$&0.6$\pm{0.2}$\\
&narrow3&\twCO~(J=1--0)&1.78$\pm{0.04}$&-43.20$\pm{0.04}$&2.39$\pm{0.09}$\\
&&\thCO~(J=1--0)&0.17$\pm{0.03}$&-43.2$\pm{0.2}$&2.0$\pm{0.4}$\\
&broad&\twCO~(J=1--0)&0.18$\pm{0.02}$&-36.6$\pm{0.9}$&20$\pm{2}$\\
&&\thCO~(J=1--0)&$\leq{0.06}$\tablenotemark{[1]}&-&-\\
\tableline
P1\tablenotemark{[2]}&narrow&\twCO~(J=1--0)&5.7$\pm{0.2}$&-43.29$\pm{0.02}$&1.16$\pm{0.05}$\\
&&\thCO~(J=1--0)&0.66$\pm{0.10}$&-43.64$\pm{0.10}$&1.4$\pm{0.3}$\\
\tableline
P2\tablenotemark{[2]}&narrow&\twCO~(J=1--0)&2.0$\pm{0.3}$&-43.46$\pm{0.05}$&0.9$\pm{0.2}$\\
&&\thCO~(J=1--0)&$\leq{0.19}$\tablenotemark{[1]}&-&-\\
&broad&\twCO~(J=1--0)&2.13$\pm{0.08}$&-39.5$\pm{0.2}$&6.8$\pm{0.4}$\\
&&\thCO~(J=1--0)&$\leq{0.19}$\tablenotemark{[1]}&-&-\\
\tableline
P3\tablenotemark{[2]}&broad&\twCO~(J=1--0)&1.36$\pm{0.07}$&-29.7$\pm{0.3}$&11.3$\pm{0.7}$\\
&&\thCO~(J=1--0)&$\leq{0.20}$\tablenotemark{[1]}&-&-\\
\tableline
P4\tablenotemark{[2]}&broad&\twCO~(J=1--0)&1.14$\pm{0.07}$&-35.9$\pm{0.3}$&8.3$\pm{0.6}$\\
&&\thCO~(J=1--0)&$\leq{0.20}$\tablenotemark{[1]}&-&-\\
\tableline\tableline
\end{tabular}
\tablenotetext{[1]}{No \thCO\ emission visible, where we use the value of RMS as an upper limit.}
\tablenotetext{[2]}{A single point, selected as representative one (shown in Figure~\ref{f:regdspec}).}
\end{center}
\end{table*}

We also detect broad \twCO~(J=1--0) emission lines inside the bright radio shell of the remnant.
Broad CO emission lines are detected in the north-western part in region~D (see Figure~\ref{f:regdspec}).
There are three narrow components in this region, around $-$51.5~\km\ps, $-$45~\km\ps, and $-$43~\km\ps\ (see Table~\ref{tab:fitparad}). 
The line center of broad component is around $-36$~\km\ps, which is far from the $-$51.5~\km\ps\ component. Considering the velocity and position close to that of $-$43~\km\ps\ MC, the broad component is associated to the $-$43~\km\ps\ MC.
Note that there is no broad component associated with the $-$51.5~\km\ps\ component which corresponds to the binary molecular clump.
We have not found any evidence of the binary clump being shocked by the remnant.
We choose four representative points with a strong narrow emission line at $-$43~\km\ps\ and broad emission lines at other three different velocities for spectral analysis, namely P1, P2, P3, and P4 (shown in Figure~\ref{f:regdspec}).
By spectral analysis of the emission lines at these four points, the physical states of quiet and shocked molecular gases in region~D are studied.
The fitting results are listed in Table~\ref{tab:fitparad}.
In region~D, the velocity shift of the broad component becomes larger as it goes further inside the projected extent of the remnant.
At the positions of P3 and P4, we could not see the associated narrow component but only the broad components.

There is significant mid-IR emission (12~$\mu$m and 22~$\mu$m) in the south-eastern part of region~D (see Section~\ref{subsec:morphology}), which could be emitted by polycyclic aromatic hydrocarbon (PAH), hot dust, or both.
The source of mid-IR emission is not clear.
The morphology of mid-IR emission has two peaks similar as the $-51.5$~\km\ps\ component, but with the position shifted to the $-45$~\km\ps\ protrusion. 
Both the molecular shock in the $-45$~\km\ps\ MC and the disturbed molecular gas in the $-51.5$~\km\ps\ binary molecular clump may contribute to the mid-IR emission.

\begin{figure}[ptbh!]
\centerline{{\hfil\hfil
\psfig{figure=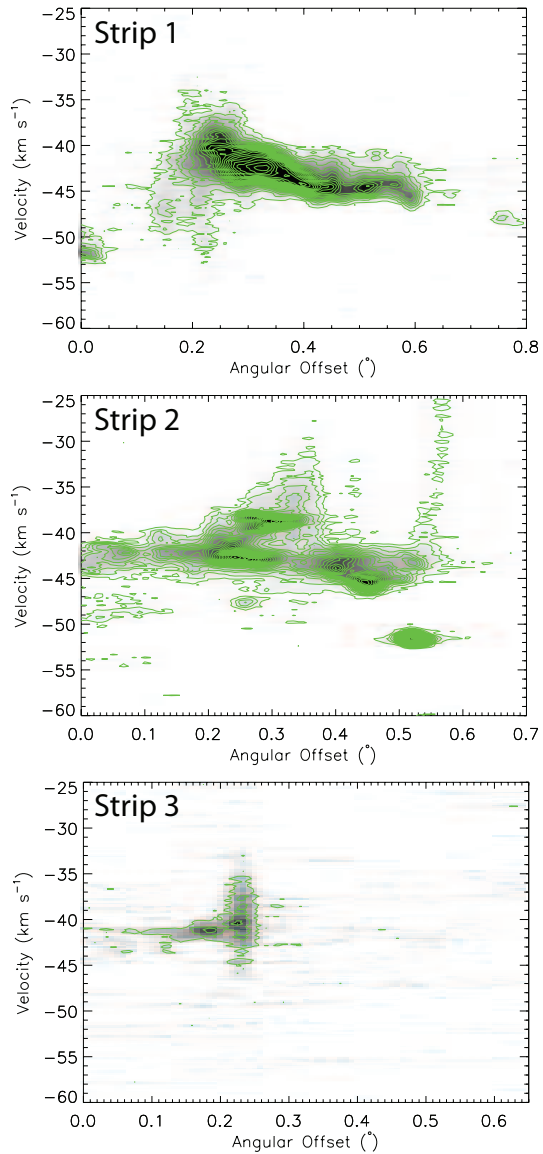,height=6.0in,angle=0, clip=}
\hfil\hfil}}
\caption{
Position-velocity maps of \twCO~(J=1--0) emission along the strips indicated by the green rectangular regions in Figure~\ref{f:tricolor}, with the directions indicated by the attached arrows. The contour levels are from $3\sigma$ and in a step of $1\sigma$. The RMS is $\sim$0.5~K. 
}
\label{f:pvmap}
\end{figure}

In Figure~\ref{f:pvmap}, we present the position-velocity distributions of \twCO~(J=1--0) emission along the strips indicated by the green rectangular regions shown in Figure~\ref{f:tricolor}, which are perpendicular to the radio continuum shell of the remnant.
The broad CO wings are prominent and near the border of the remnant in strip~1 (region~A1) and strip~3 (region~C). 
In strip~1, the emission peak is inside the projected extent of the remnant, which could be due to the effect of line-of-sight superimposition. 
In strip~3, the intensity of broadened component is comparable to that of narrow component, and both are weak.
It indicates that the amount of molecular gas in this region is not as large as in the other regions; nevertheless, the percentage of shocked molecular gas is higher than in the other regions.

In strip~2, the broad CO wings are presented at two positions, with the angular offsets of 0.35$^\circ$ and 0.55$^\circ$ (see Figure~\ref{f:pvmap}). One is at the radio peak (region~B1), and the other is further inside the projected extent of the remnant (region~D). 
Note that region~B1 is in the middle of the conjunction area between W3 and \snr, where CO emission protrudes from the W3 region into the \snr\ region.
In this area, the radio continuum emission and the broad CO wings are intense, which indicates a strong interaction between the remnant and the MCs. 
Region~D appears further inside the projected extent of the remnant than the other regions, and its broad component also has the largest line-width.

We confirm, based on both morphological and dynamical evidences, the association between the remnant and the $\sim$$-$45~\km\ps\ MC, which is itself associated with the W3 H~{\sc ii} complex.
Therefore, \snr\ is at the same distance as the W3/W4 complex, which is $1.95\pm0.04$~kpc. Accordingly, the physical size of the remnant is $65\times48$~pc$^{2}$. 

\section{Discussion}\label{sec:discuss}
\subsection{Physical Conditions of the Molecular Gas}\label{subsec:phypar}
\begin{table*}\footnotesize
\begin{center}
\caption{Derived physical parameters.\label{tab:phypara}}
\begin{tabular}{llcccccc}
\tableline\tableline
region&component&$T_{\rm ex}$\tablenotemark{[1]}&$\tau$(\thCO)\tablenotemark{[1]}&\ncol\tablenotemark{[2]}&$M$\tablenotemark{[2]}&$M_{\rm vir}$\tablenotemark{[3]}\\
&&(K)&&(10$^{20}$ cm$^{-2}$)&($\Msun$)&($\Msun$)\\
\tableline
A1&narrow&$>$4.3&$<0.2$&$<9.7$ (8.8)&$<22{d}^2_{1.95}$ (63$d^2_{1.95}$)&5.1$\E{3}$\du \\
&broad&$>$4.0&$<0.3$&$<45$ (28)&$<1.1\E{2}{d}^2_{1.95}$ (2.0$\E{2}d^2_{1.95}$)&8.3$\E{4}$\du \\
\tableline
A2&narrow&8.1&0.8&31 (19)&$1.2\E{2}{d}^2_{1.95}$ ($84d^2_{1.95}$) &6.0$\E{2}$\du  \\
&broad&$\ge$5.9&$\le$0.3&51 (45)&59${d}^2_{1.95}$ (2.0$\E{2}d^2_{1.95}$) &1.1$\E{4}$\du \\
\tableline
B1&narrow1&6.8&0.3&18 (12)&1.6$\E{2}{d}^2_{1.95}$ (2.1$\E{2}d^2_{1.95}$) &1.2$\E{3}$\du  \\
&narrow2&7.5&0.3&13 (9.2)&$1.0\E{2}{d}^2_{1.95}$ (1.6$\E{2}d^2_{1.95}$) &4.6$\E{2}$\du  \\
&broad&$>$5.0&$<0.1$&$<18 $(38)&$<79{d}^2_{1.95}$ (6.4$\E{2}d^2_{1.95}$) &3.9$\E{4}$\du \\
\tableline
B2&narrow1&8.9&0.3&9.3 (11)&7.2${d}^2_{1.95}$ (17$d^2_{1.95}$) &1.1$\E{2}$\du  \\
&narrow2&6.8&0.4&15 (26)&8.0${d}^2_{1.95}$ (41$d^2_{1.95}$) &1.6$\E{3}$\du  \\
&broad&$>$4.3&$<0.3$&$<18 $(11)&$<6.8{d}^2_{1.95}$ (18$d^2_{1.95}$) &2.4$\E{3}$\du \\
\tableline
C&narrow&$>$3.3&$<1.1$&$<21$ (1.7)&$<42{d}^2_{1.95}$ (14$d^2_{1.95}$) &9.7$\E{2}$\du  \\
&broad&$>$3.8&$<0.5$&$<50$ (15)&$<1.0\E{2}{d}^2_{1.95}$ (1.2$\E{2}d^2_{1.95}$) &2.3$\E{4}$\du \\
\tableline
\multicolumn{7}{c}{Example points in region~D}\\
\tableline
P1&narrow&9.0&0.1&6.8 (13)&0.9${d}^2_{1.95}$ (1.6$d^2_{1.95}$) &40\du  \\
\tableline
P2&narrow&$>$5.1&$<0.5$&$<7.1$ (3.4)&$<0.2{d}^2_{1.95}$ (0.4$d^2_{1.95}$) &24\du  \\
&broad&$>$5.3&$<0.4$&$<51$ (28)&$<1.6{d}^2_{1.95}$ (3.6$d^2_{1.95}$) &1.4$\E{3}$\du  \\
\tableline
P3&broad&$>$4.4&$<0.9$&$<1.3\E{2}$ (29)&$<4.3{d}^2_{1.95}$ (3.8$d^2_{1.95}$) &3.8$\E{3}$\du  \\
\tableline
P4&broad&$>$4.1&$<1.2$&$<1.2\E{2}$ (18)&$<4.0{d}^2_{1.95}$ (2.3$d^2_{1.95}$) &2.1$\E{3}$\du  \\
\tableline\tableline
\end{tabular}
\tablenotetext{[1]}{Using the assumption of local thermal equilibrium (LTE). For components with no \thCO~(J=1--0) emission detected, we use the values of RMS as its upper limit. See the details of calculation method in Section~\ref{subsec:phypar}.}
\tablenotetext{[2]}{Derived from \thCO\ column density by assuming the \thCO\ abundance of 1.4$\E{-6}$ \citep{Ripple+2013}. For comparison, we also show the values in the brackets, which are estimated by using the conversion factor ${N}({\rm H}_2)/{W}(^{12}{\rm CO)}\simeq1.8\times10^{20}~{\rm cm}^{-2}~{\rm K}^{-1}~{\rm km}^{-1}$~s \citep{Dame+2001}. 
}
\tablenotetext{[3]}{Calculated by ${k}_{2}\times{L}\times\Delta v^{2}$, where $k_2$ is 105 \citep{MacLaren+1988}, $L$ is the size of the region, and $\Delta v$ is the velocity width (FWHM) of \twCO~(J=1--0).}
\end{center}
\end{table*}

We estimate the physical parameters of molecular gas in the selected regions.
For \twCO~(J=1--0), given the background temperature ${T}_{\rm bg}=2.73$~K, we get the excitation temperature as:
\begin{equation}
T_{\rm ex}=5.53 ({\rm log}(1+\frac{5.53}{T_{\rm mb,peak,^{12}CO}/(1-{\rm exp}(-\tau_{\rm ^{12}CO}))+0.84}))^{-1}~({\rm K}),
\label{eq:tex}
\end{equation}
where $\tau$ and ${T}_{\rm mb,peak}$ are the optical depth and the peak ${T}_{\rm mb}$ of the corresponding emission lines, respectively.
In the assumption of local thermodynamic equilibrium (LTE), the excitation temperatures of \twCO~(J=1--0) and \thCO~(J=1--0) should be the same;
we could derive the optical depth of \thCO~(J=1--0) as:
\begin{equation}
\tau_{\rm ^{13}CO}=-{\rm log}(1-\frac{T_{\rm mb,peak,^{13}CO}/f_{\rm ^{13}CO}}{5.29(\frac{1}{{\rm exp}(5.29/T_{\rm ex})}-0.17)}),
\label{eq:tau}
\end{equation}
and the column density of \thCO\ as \citep[e.g.][]{Garden+1991}:
\begin{equation}
\begin{split}
N_{\rm ^{13}CO}&=2.50\E{14} \frac{T_{\rm ex}+0.88}{1-{\rm exp}(-5.29/T_{\rm ex})} \int \tau_{\rm ^{13}CO} dv~({\rm cm}^{-2}) \\
 &\simeq2.50\E{14} \frac{T_{\rm ex}+0.88}{1-{\rm exp}(-5.29/T_{\rm ex})} \tau_{\rm ^{13}CO} \Delta v_{\rm ^{13}CO},
\end{split}
\label{eq:ncol}
\end{equation}
where $f_{\rm ^{13}{CO}}$ and $\Delta v_{\rm ^{13}{CO}}$ are the area beam-filling factor and the full width at half maximum (FWHM) of \thCO~(J=1--0), respectively. 
The area beam-filling factor of \twCO~(J=1--0) is assumed to be unity in our calculation, which is reasonable considering that the selected regions are filled with \twCO~(J=1--0) emission. 
However, \thCO~(J=1--0) is not in this case, and its area beam-filling factor is estimated by the ratio of the number of points with detected \thCO\ emission to that with detected \twCO\ emission. 
For the regions without detected \thCO~(J=1--0), the area beam-filling factor of \thCO\ is assumed to be the minimum one of that from the regions with detected \thCO~(J=1--0), which is $1/4$.

Applying $\tau_{\rm ^{12}CO}/\tau_{\rm ^{13}CO}\simeq N_{\rm ^{12}CO}/N_{\rm ^{13}CO}\simeq [^{12}{\rm C}/^{13}{\rm C}]=70$ \citep{Milam+2005} to Equation~(\ref{eq:tex}) and Equation~(\ref{eq:tau}), we calculated the $T_{\rm ex}$ and $\tau_{\rm ^{13}CO}$ recursively. 
The derived physical parameters are listed in Table~\ref{tab:phypara}.
For the MC with a low column density, the photodissociation rates can be different for \twCO\ and \thCO, which may cause $N_{\rm ^{12}CO}/N_{\rm ^{13}CO}\gtrsim [^{12}{\rm C}/^{13}{\rm C}]$. But this effect only causes an increase of the ratio up to 25 percent \citep{Szuecs+2014}.

Most of the broad components do not have corresponding \thCO~(J=1--0) emission detected, which indicates that their \twCO~(J=1--0) emissions are optically thin.
Therefore, the excitation temperature of these broad components cannot be well determined.
We detect broad \thCO~(J=1--0) emission in region~A2, and the ratio of brightness temperatures $f_{\rm ^{13}CO}\times T_{\rm mb,peak,^{12}CO}/T_{\rm mb,peak,^{13}CO}$ are $4\pm2$ for the broad component and $1.88\pm0.02$ for the narrow component. 
Considering the possibility of this broad \twCO~(J=1--0) emission being optically thin too, we give the derived excitation temperature as a lower limit in region~A2 either.
Note that, the large velocity widths of the lines support high excitation temperature.

The stability of MC could be investigated by the virial theorem. The mass of all the molecular clouds is smaller than their virial mass (see Table~\ref{tab:phypara}). However, it can be caused by the small sizes of the selected regions, which are about one order of magnitude smaller than the size of the target clouds. Taking this into account, the narrow components should be stable, with gravity and disturbance in equilibrium. 
Since the broad components are mainly distributed in small regions, no correction for size is needed. 
The mass of the broad components is at least two orders of magnitude lower than their virial mass, which indicates the existence of strong perturbations in these regions.

\subsection{Properties of SNR \snr}\label{sec:snr}

Adopted from the radio continuum extent of the remnant ($\sim90'$, see Figure~\ref{f:tricolor}), the radius of the remnant is $r_{\rm s}\sim25.5$\du~pc.
Assuming the remnant is in the Sedov phase, and applying the velocity of the remnant's shock $v_{\rm s}=340\pm{37}$~\km\ps\ and the remnant's ambient particle density $n_{\rm 0}=0.32\pm0.10$~~cm$^{-3}$ \citep[derived from the X-ray study by][]{LazendicSlane2006}, we get the age of the remnant as 
$t\sim (2.9\pm{0.3})\E{4}$\du~yr, and the explosion energy as
$E\sim(1.3\pm{0.8})\E{51} {d}_{1.95}^{3}$~erg.
Alternatively, the remnant may already enter the radiative phase. In this case, the age of the remnant is $t=(2 r_{\rm s})/(7 v_{\rm s})\sim (2.1\pm{0.2})\E{4}$\du~yr \citep{McKeeOstriker1977}, 
and the explosion energy is 
$E=6.8\E{43} n_{\rm 0}^{1.16} (\frac{v_{\rm s}}{1~{\rm km}~{\rm s}^{-1}})^{1.35} (\frac{r_{\rm s}}{1~{\rm pc}})^{3.16} \zeta_{\rm m}^{0.161}$
$\sim(1.6\pm{0.9})\E{51} {d}_{1.95}^{3.16}$~erg,
where $\zeta_{\rm m}=Z/Z_{\rm \odot}=1$ \citep{Cioffi+1988}.
The derived energies have large errors; however, they are basically consistent in the two cases. 

We adopt the same velocity of the remnant's shock and ambient particle density as that in \cite{LazendicSlane2006}. Nevertheless, we use a revised distance of 1.95~kpc, which is smaller than that of 2.2~kpc used in \cite{LazendicSlane2006}. Therefore, in the case of the Sedov phase, we get the age of the remnant very similar to that from \cite{LazendicSlane2006}, and get the explosion energy more different, since the age is proportional to the distance, whereas the explosion energy is proportional to the 3rd power of the distance.

\snr\ is surrounded by a partial molecular shell from the east to the southwest (see panel ({\it a}) in Figure~\ref{f:stamp12} and Figure~\ref{f:tricolor}).
This partial molecular shell could be swept up by either the SNR or the stellar wind of the SNR's progenitor.
If it was swept up by the progenitor's stellar wind, the progenitor's mass could be estimated by using the linear relationship between the size of the wind-blown bubble in a molecular environment and the star's initial mass \citep{Chen+2013}.
The radius of the wind-blown bubble is no less than the radius of the partial molecular shell, which is $\gtrsim$$25$~pc. Then the mass of \snr's progenitor is $\gtrsim$$28~\Msun$.

\subsection{Binary molecular clump}\label{subsec:bc}
\begin{table*}\footnotesize
\begin{center}
\caption{Fitted and derived parameters of thermal molecular lines from the binary molecular clump.\label{tab:bc}}
\begin{tabular}{llcccccc}
\tableline\tableline
region\tablenotemark{[1]}&Line&Peak \tmb &Center \vlsr&FWHM\\
&&(K)&(\km\ps)&(\km\ps)\\
\tableline
bcu&\twCO~(J=1--0)&7.85$\pm{0.08}$&-51.704$\pm{0.005}$&0.96$\pm{0.01}$\\
&\thCO~(J=1--0)&1.28$\pm{0.05}$&-51.89$\pm{0.02}$&0.81$\pm{0.03}$\\
\tableline
bcl&\twCO~(J=1--0)&5.95$\pm{0.07}$&-51.411$\pm{0.006}$&1.03$\pm{0.02}$\\
&\thCO~(J=1--0)&0.73$\pm{0.05}$&-51.58$\pm{0.03}$&0.68$\pm{0.06}$\\
\tableline
bct&\twCO~(J=1--0)&1.88$\pm{0.07}$&-50.73$\pm{0.02}$&1.12$\pm{0.05}$\\
&\thCO~(J=1--0)&0$\pm{0.07}$&-&-\\
\tableline\tableline
\multicolumn{6}{c}{Derived parameters\tablenotemark{[2]}}\\
\tableline
region\tablenotemark{[1]}&$T_{\rm ex}$&$\tau$(\thCO)&\ncol&$M$&$M_{\rm vir}$\\
&(K)&&(10$^{20}$ cm$^{-2}$)&($\Msun$)&($\Msun$)\\
\tableline
bcu&11.2&0.2&11 (14)&59${d}^2_{1.95}$ ($1.0\E{2}d^2_{1.95}$) &2.5$\E{2}$\du \\
bcl&9.3&0.2&5.6 (12)&16${d}^2_{1.95}$ (48$d^2_{1.95}$) &2.5$\E{2}$\du  \\
bct&$>$5.0&$<$0.2&$<$2.9 (4.0)&$<$1.8${d}^2_{1.95}$ (10$d^2_{1.95}$) &1.9$\E{2}$\du  \\
\tableline\tableline
\end{tabular}
\tablenotetext{[1]}{We divided the binary system into three regions for spectral analysis, bcu for the upper clump, bcl for the lower clump, and bct for the tail of the binary clump (see Figure~\ref{f:bcir}).}
\tablenotetext{[2]}{Use the same calculation method as in Table~\ref{tab:phypara}.}
\end{center}
\end{table*}

We investigate the physical state of the binary molecular clump by analyzing its CO emission lines. 
The CO spectra are extracted from three regions of the binary clump, namely, bcu, bcl, and bct, corresponding to the upper clump, the lower clump, and the tail of the binary clump, respectively (see Figure~\ref{f:bcir}).
Using the same method as in Section~\ref{subsec:phypar}, we estimate the physical parameters of the molecular gas in these regions.
The fitted and derived parameters of the CO emission lines are listed in Table~\ref{tab:bc}.

The mass of the upper clump is smaller than its virial mass (the virial parameter $\alpha_{\rm vir,bcu}=M_{\rm vir,bcu}/M_{\rm bcu}\sim4$). 
However, the factor is only a few, indicating that the upper clump is about stable.
For the lower clump, the mass is about one order of magnitude smaller than its virial mass ($\alpha_{\rm vir,bcl}\sim16$). Therefore, the lower clump is significantly disturbed.
The virial parameter of the tail is $\alpha_{\rm vir,bct}\gtrsim1\E{2}$, implying that the tail is loosely bound by the system.

The clumps in a binary system are affected by tidal force, which could destroy these clumps. 
To estimate the effect of tidal force on the upper clump, we derive the ratio between the tidal force and the self-gravity as
$\gamma_{\rm bcu}\sim 4 M_{\rm bcl} M_{\rm bcu}^{-1} r r_{\rm bcu}^{3} (r^2-r_{\rm bcu}^2)^{-2}$ 
$\sim 0.82 (1.90~{\rm sin}^{-1.5}\theta-0.67~{\rm sin}^{0.5}\theta)^{-2}$
$\lesssim0.54$, 
where $M_{\rm bcl}$ and $M_{\rm bcu}$ are the mass of the lower and upper clumps, respectively, $r$ is the distance between the two clumps, and $r_{\rm bcu}$ is the radius of the upper clump. 
For the lower clump, we get the tidal force factor $\gamma_{\rm bcl}\lesssim1.38$. 
Therefore, the tidal force could not directly destroy the upper clump in the binary system, but could destroy the lower clump.
In any case, the tidal disturbance will play an important role during the evolution of the binary system.
The tail structure here in the binary system resembles the tidal tails of interacting galaxies, e.g. NGC~3256 and NGC~5752/4 \citep{KeelBorne2003, Trancho+2007, Smith+2010}, etc.
The loosely bound tail material could get stripped during the interaction process. 
The angular momentum would be taken away, and hence the binary system becomes stable.
Such molecular clump interaction may induce star-formations;
nevertheless, the clumps could be stripped and lose their mass.

\subsection{Star Formation Activity}\label{subsec:SF}
\begin{figure*}[ptbh!]
\centerline{{\hfil\hfil
\psfig{figure=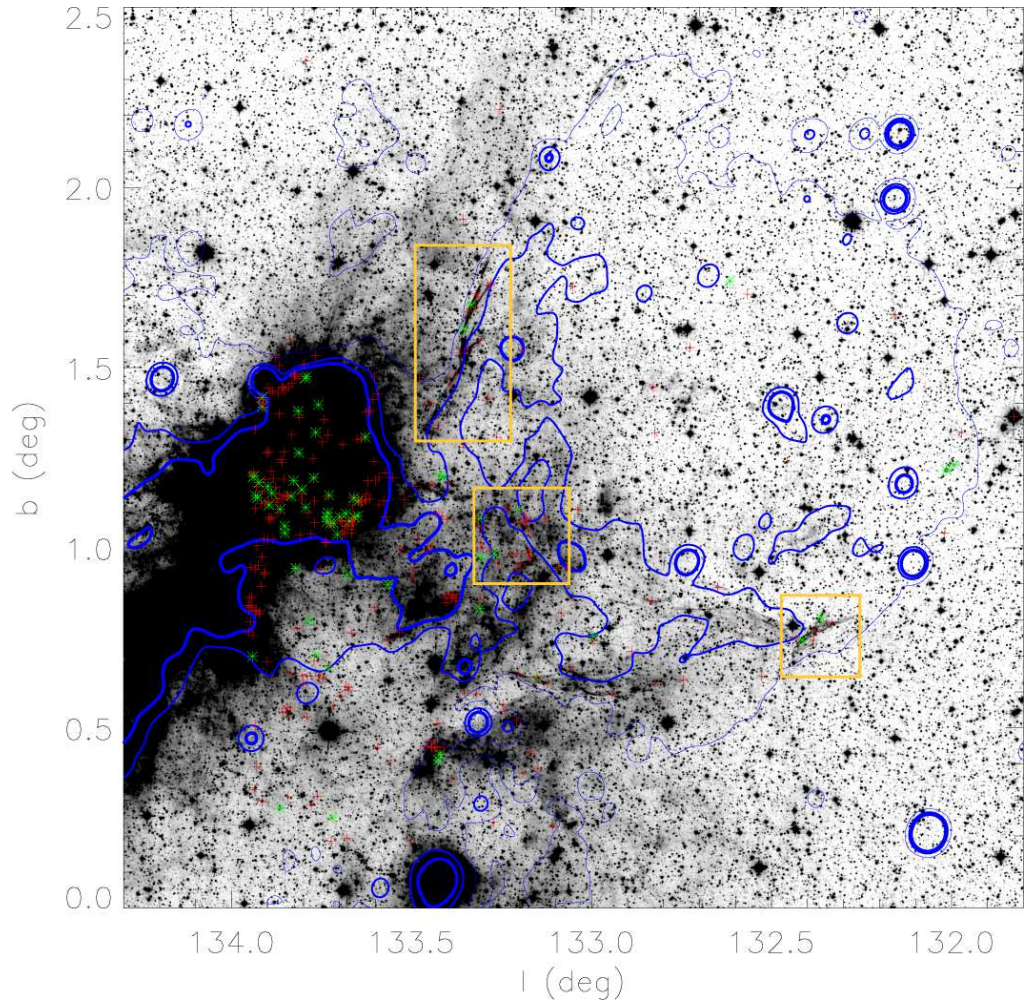,height=5.2in,angle=0, clip=}
\hfil\hfil}}
\caption{
Image of 4.6~$\mu$m emission observed by WISE, overlaid with the same 408~MHz continuum contours as in Figure~\ref{f:stamp12} (in blue). 
The positions of young stellar object candidates (YSOc) are marked with green stars (type Class~I) and red pluses (type Class~II). 
Aggregated YSOc in the \snr\ region are divided into three clusters, which are marked with yellow rectangular boxes in the east, southeast, and southwest of the remnant.
}
\label{f:yso}
\end{figure*}

We have selected young stellar object candidates (YSOc) from the IR data (see Figure~\ref{f:yso}). 
Using the color-color and {\it WISE} photometry criteria described in \cite{Koenig+2012}, disk-bearing young stars are identified, of which the IR colors are distinctly different from those of diskless objects. 
Diskless young stars cannot be distinguished from unrelated field objects based on IR colors alone. 
For the sources that are not detected in the {\it WISE} [12] band, the $K_{\rm s}$$-$[3.4] versus [3.4]$-$[4.6] color--color diagrams are constructed based on their dereddened photometry in the {\it WISE} [3.4] and [4.6] bands, in combination with the dereddened 2MASS $K_{\rm s}$ photometry.
To deredden the photometry, we estimate the extinction by the locations in the $J$$-$$H$ versus $H$$-$$K_{\rm s}$ color--color diagram \citep[see details in][]{Fang+2013}. 
We have further checked the YSOc by eye to exclude the sources that are not point-like, to maximally eliminate the contamination from small shocked clumps.

The YSOc are mainly distributed within the W3 region, and also in the conjunction area between W3 and \snr\ (see Figure~\ref{f:yso}).
Most of YSOc in the \snr\ region are aggregated and along the outer rim of the radio shell of \snr.
We could generally divide them into three clusters.
The most distinctive one is in the east of the remnant, which is spatially corresponding not only with 4.6~$\mu$m filaments but also with the eastern edge of \snr\ where radio emission is the most steep (see Figure~\ref{f:yso}).
In this region, we detect weak CO emissions, with line wing broadening features.
Another cluster is in the southwest of the remnant, around the top-end of the remnant's radio shell, where the border of radio continuum emission is a little dent. It is also associated with a shorter curved 4.6~$\mu$m filament.
The rests of aggregated YSOc are in the southeast of the remnant, around the conjunction point between the remnant and the W3 region, where both the radio continuum emission and the \twCO~(J=1--0) emission are strong.
The distribution of aggregated YSOc shows very well morphological correlations with the radio emission of \snr.
It provides a strong evidence of association between the SNR and the underlying star-formation activities, which indicates the ignition of star-formation at the periphery of a well-developed star-forming region.
Due to the short lifetime of the remnant, the related star-formation activities are hardly being triggered by \snr\ directly, but are probably triggered by the stellar wind of \snr's progenitor. 
It indicates that the partial molecular shell surrounding the remnant from the east to the southwest was swept up by the progenitor's stellar wind. 
Therefore, the mass of \snr's progenitor is $\gtrsim$$28~\Msun$ (see Section~\ref{sec:snr}).

\cite{Oey+2005} suggested that W3/W4 was a three-generation hierarchical star-forming system.
The current star-forming activities in W3 could be triggered by the OB association IC~1795.
The age of IC~1795 is 3 to 5~Myr \citep{Oey+2005}, which could belong to the same generation as \snr's progenitor. The distance between \snr's geometrical center and IC~1795 is about 41\du~pc. 
If \snr's progenitor were a runaway star from IC~1795, its velocity would be about 8 to 13\du~\km\ps\ that is moderate \citep[e.g.][]{Banerjee+2012}. 
Indeed, the O- and B-type stars are widely dispersed across the W3 complex \citep{Kiminki+2015}.
It is possible that \snr's progenitor used to be in IC~1795. 
This suggests that the propagation of star-formation could be very fast, and the case of the next generation triggered star-formation could be truly complicated.

\section{Conclusions}\label{sec:conclusion}

We present millimeter observations in CO emission lines toward \snr. 
Substantial molecular gas around $-45$~\km\ps\ is detected in the conjunction region between the SNR~\snr\ and the nearby H~{\sc ii} region/MC complex W3.
This molecular gas is distributed along the radio continuum shell of the remnant.
Furthermore, the shocked molecular gas indicated by line wing broadening features is also distributed along the radio shell and inside it. By both morphological correspondence and dynamical evidence, we confirm that the SNR~HB~3 is interacting with the $-45$~\km\ps\ MC, in essence, with the nearby H~{\sc ii} region/MC complex W3.
The red-shifted line wing broadening features indicate that the remnant is at the nearside of the MC.
With this association, we could place the remnant at the same distance as the W3/W4 complex, which is $1.95\pm0.04$~kpc.
We also find a spatial correlation between the aggregated YSOc and the shocked molecular strip which is associated with the remnant.

Particularly, a binary clump at ($l=132^{\circ}.94, b=1^{\circ}.12$) around $-51.5$~\km\ps\ inside the remnant's radio shell has been found, and it is associated with significant mid-IR emission. 
The binary system also has a tail structure resembling the tidal tails of interacting galaxies.
According to the analysis of CO emission lines, the larger clump in this binary system is approaching stability, and the smaller clump is significantly disturbed.

\acknowledgments
We are grateful to all the members in the Milky Way Scroll Painting-CO line survey group, especially the staff of Qinghai Radio Observing Station at Delingha for the support during the observation.
We thank the anonymous referee for providing very helpful comments that improved the paper and its conclusions.
This work is supported by NSFC grants 11233007, 11233001, and 11403104, and Jiangsu Provincial Natural Science Foundation grant BK20141044.
Y.C. acknowledges support by 973 Program grant 2015CB857100 and grant 20120091110048 from the Educational Ministry of China.
This work is also supported by the Strategic Priority Research Program of the Chinese Academy of Sciences, grant No.\ XDB09000000.
The research presented in this paper has used data from the Canadian Galactic Plane Survey, a Canadian project with international partners, supported by the Natural Sciences and Engineering Research Council.
This publication makes use of data products from the Two Micron All Sky Survey, which is a joint project of the University of Massachusetts and the Infrared Processing and Analysis Center/California Institute of Technology, funded by the National Aeronautics and Space Administration and the National Science Foundation.
This publication makes use of data products from the Wide-field Infrared Survey Explorer, which is a joint project of the University of California, Los Angeles, and the Jet Propulsion Laboratory/California Institute of Technology, funded by the National Aeronautics and Space Administration.

\bibliographystyle{apj}
\bibliography{ms.bbl}

\end{CJK*}
\end{document}